\documentclass[aps,twocolumn,amsmath,amssymb,prm,showpacs]{revtex4-1}

\usepackage{graphicx}
\usepackage{comment}
\usepackage[normalem]{ulem}
\usepackage{color}
\usepackage[colorlinks=true,linkcolor=blue,anchorcolor=blue,citecolor=blue]{hyperref}%
\usepackage{amsmath}
\usepackage{soul}

\definecolor{darkpastelgreen}{rgb}{0.01, 0.75, 0.24}

\makeatletter
\def\blfootnote{\xdef\@thefnmark{}\@footnotetext}
\makeatother

\begin{document}
%
%
\title{Thermodynamic evidence of fermionic behavior in the vicinity of one-ninth plateau in a kagome antiferromagnet}

\author{Guoxin Zheng$^1$, Dechen Zhang$^{1}$, Yuan Zhu$^{1}$, Kuan-Wen Chen$^1$,  Aaron Chan$^{1}$, Kaila Jenkins$^{1}$, Byungmin Kang$^2$, Zhenyuan Zeng$^{3,4}$, Aini Xu$^{3,4}$, D. Ratkovski$^{5}$, Joanna Blawat$^{6}$, Alimamy F. Bangura$^{5}$, John Singleton$^{6}$, Patrick A. Lee$^2$, Shiliang Li$^{3,4,7}$ }
\author{Lu Li$^{1}$}
\email{luli@umich.edu}

\affiliation{
	$^1$Department of Physics, University of Michigan, Ann Arbor, MI 48109, USA\\
	$^2$Department of Physics, Massachusetts Institute of Technology, Cambridge, MA 02139, USA\\
	$^3$Beijing National Laboratory for Condensed Matter Physics, Institute of Physics, Chinese Academy of Sciences, Beijing 100190, China\\
	$^4$School of Physical Sciences, University of Chinese Academy of Sciences, Beijing, 100190, China\\
	$^5$National High Magnetic Field Laboratory, 1800 East Paul Dirac Drive, Tallahassee, Florida 32310-3706, USA\\
	$^6$National High Magnetic Field Laboratory, MS E536, Los Alamos National Laboratory, Los Alamos, NM 87545, USA\\
	$^7$Songshan Lake Materials Laboratory, Dongguan, Guangdong, 523808, China
}

\date{\today}
\begin{abstract}
	The spin-1/2 kagome Heisenberg antiferromagnets are believed to host exotic quantum entangled states. Recently, the report of 1/9 magnetization plateau and magnetic oscillations in a kagome antiferromagnet YCu$_3$(OH)$_6$Br$_2$[Br$_x$(OH)$_{1-x}$] (YCOB) have made this material a promising candidate for experimentally realizing quantum spin liquid states. Here we present measurements of the specific heat $C_p$ in YCOB in high magnetic fields (up to 41.5 Tesla) down to 0.46 Kelvin, and the 1/9 plateau feature has been confirmed. Moreover, the temperature dependence of $C_p/T$ in the vicinity of 1/9 plateau region can be fitted by a linear in $T$ term which indicates the presence of a Dirac spectrum, together with a constant term, which indicates a finite density of states (DOS) contributed by other Fermi surfaces. Surprisingly the constant term is highly anisotropic in the direction of the magnetic field. Additionally, we observe a double-peak feature near $30$~T above the 1/9 plateau which is another hallmark of fermionic excitations in the specific heat.
\end{abstract}


\maketitle                   
\renewcommand{\thesubsection}{\Alph{subsection}}
\renewcommand{\thesubsubsection}{\textit{\alph{subsubsection}}}

\section{Introduction}
Quantum spin liquids (QSLs) have played an essential role in condensed matter physics since Anderson proposed the resonating-valence-bond (RVB) model in 1973 \cite{Anderson1973}. The spin-1/2 kagome Heisenberg antiferromagnet (KHA) exhibits a high degree of geometric frustration and is one of the most promising candidates for hosting QSLs \cite{Savary2016, Zhou2017, Norman2016}. Theoretically, the presence of QSL on the KHA has been confirmed by density matrix renormalization group (DMRG) simulations \cite{Yan2011}, but its precise ground state remains an open question, with two main possibilities: the gapped $Z_2$ spin liquid \cite{Yan2011, Moessner2001,Depenbrock2012}, and the gapless $U(1)$ Dirac spin liquid (DSL) \cite{Ran2007, Iqbal2013, Hastings, He2017}. Beyond the ground state at zero field, more exotic quantum entangled states can emerge under magnetic fields, such as the unconventional 1/9 magnetization plateau, which might be described by a topological $Z_3$ QSL \cite{Nishimoto2013} or a gapless valence-bond-crystal state \cite{Fang2023}, though its nature remains elusive. A recent projected Monte Carlo study supports a Z3 spin liquid scenario with fermionic spinons. \cite{He2024}

Experimentally, the most extensively studied QSL candidate in the kagome system so far is herbertsmithite [ZnCu$_3$(OH)$_6$Cl$_2$]~\cite{Shores2005, Han2012}. The difficulty in determining its ground state arises from the substitution between Zn$^{2+}$ and the two-dimensional (2D) kagome plane formed by Cu$^{2+}$, which causes the low-energy spectrum to be dominated by impurity spins~\cite{Vries2008, Freedman2010, Huang2021}. Recently, the synthesis of the KHA YCu$_3$(OH)$_6$Br$_2$[Br$_x$(OH)$_{1-x}$] (YCOB) has addressed the site mixing issue by introducing Y$^{3+}$ ions which have a much larger atom size than Cu$^{2+}$~\cite{Chen2020}. The absence of a magnetic transition down to 50~mK in YCOB makes it a compelling QSL candidate~\cite{Chen2020, Zeng2022}, even though disorder in the exchange coupling is now present~\cite{Chen2020}.

Very recently, experimental progress has come from reports of the signature of a DSL~\cite{Zheng2023, Zeng2022, Liu2022, Zeng2024}, the 1/9 magnetization plateau~\cite{Zheng2023, Jeon2024, Suetsugu2024}, and magnetic oscillations~\cite{Zheng2023} in YCOB. The Dirac spinon behavior at zero field has been inferred from specific-heat~\cite{Zeng2022, Liu2022}, nuclear magnetic resonance (NMR)~\cite{Lu2022, Li2024}, and neutron-scattering~\cite{Zeng2024} measurements, while the 1/9 magnetization plateau has only been studied by magnetization and magnetic torque measurements so far~\cite{Zheng2023,Jeon2024,Suetsugu2024}, and even the gapped or gapless nature is still under debate. Therefore, there is an urgent need for more experimental probes to investigate the nature of the unconventional 1/9 magnetization plateau.

In this paper, we report the specific heat ($C_p$) measurements on single crystals of YCOB under magnetic fields of up to $\mu_0 H=41.5$~T, with temperatures down to $T=0.46$~K. The 1/9 plateau phase previously observed in magnetization is verified in the magnetic-field dependence of the specific heat, and its gapless nature is identified from the finite value of $C_p/T$ in the $T \rightarrow 0$ limit. Moreover, the significant quadratic $T$ dependence term in $C_p$ indicates a Dirac spinon contribution, and the observed ``double-peak" structure in $C_p$ provides further evidence of fermionic behavior.

\begin{figure}[!htb]
	
	\begin{center}
		\includegraphics[width= 0.5\textwidth ]{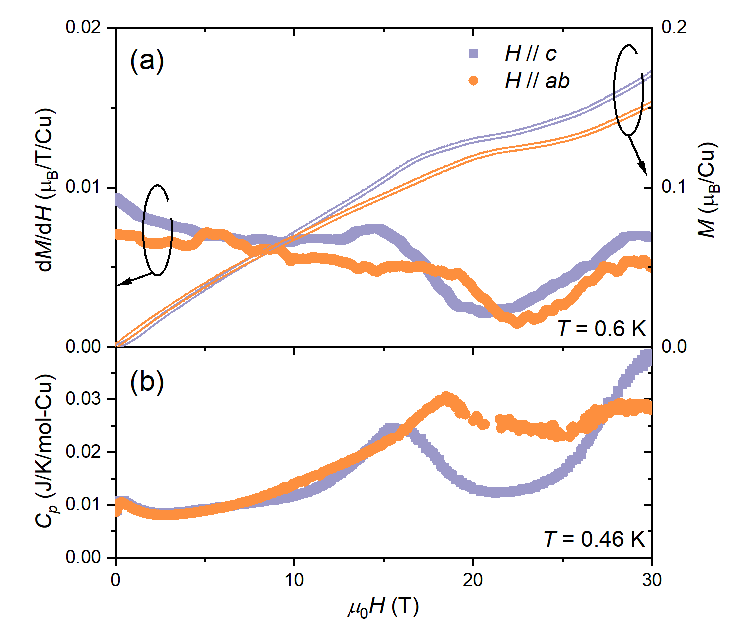}
		\caption{Magnetic field dependence of magnetization, the corresponding derivative, and specific heat around 1/9 plateau phase. (a) The thinner double lines represent the $M$ vs. $H$ data measured at 0.6 K with applied field along the $c$ (blue) and $ab$ (orange) directions. The magnetic susceptibilities $\chi_M \equiv {\rm d}M/{\rm d}H$ are plotted as the thicker dots corresponding to the vertical scale on the left side. The 1/9 magnetization plateau is observed along the $c$ axis between 15 T and 28 T and in the $ab$ plane between 20 T and 27 T. (b) The field dependence of specific heat measured at 0.46~K. The valleys centered around $\mu_0 H_0=22$~T confirmed the 1/9 plateau in magnetization.} 
		\label{fig1}
	\end{center}
\end{figure}

\section{Experiment}\label{sec_sample}
For this study, single crystals of YCOB were grown using the hydrothermal method as reported previously~\cite{Zeng2022}. The magnetization measurements on YCOB sample M1 were performed using a compensated coil spectrometer~\cite{Goddard2007, Goddard2008} in a 65~T pulsed field magnet at the National High Magnetic Field Laboratory (NHMFL), Los Alamos. The specific heat measurements on YCOB sample H1 at high fields were carried out using a membrane-based nanocalorimeter \cite{Tagliati2012} employing an ac steady-state method~\cite{Sullivan1968} in the 41.5~T Cell~6 DC field magnet at NHMFL, Tallahassee. The specific heat measurements on YCOB sample H3 at 0~T in the inset of Fig. \ref{fig2}(a) were conducted in a Quantum Design Physical Property Measurement System (PPMS) using the He-3 option.

\section{Results}
\subsection{One-ninth plateau in magnetization and specific heat}
The magnetic field ($H$) dependence of magnetization $M$ and the corresponding derivative $\chi_M \equiv {\rm d}M/{\rm d}H$ for $H \parallel c$ and $H \parallel ab$ at temperature $T=0.6$~K are shown in Fig. \ref{fig1}(a). The experimental details and sample growth information are given in Section \ref{sec_sample}. The plateau region can be characterized by the width of the valley in $\chi_M$, which spans from 15.0 T to 28.4 T when $H \parallel c$ and from 19.1 T to 27.3 T when $H \parallel ab$. This observation is consistent with the results reported in~\cite{Zheng2023, Jeon2024, Suetsugu2024}. The slightly larger 1/9 magnetization value when $H \parallel c$ can be understood by the anisotropy of the $g$-factor~\cite{Liu2022}. 
To confirm the 1/9 plateau feature and shed light on this unconventional state, we conducted specific heat measurements at high fields. The field dependence around the 1/9 plateau region is shown in Fig. \ref{fig1}(b) with applied field along $H \parallel c$ and $H \parallel ab$. The 1/9 plateau phase is visible as a dip in specific heat within the same field range. The similar behavior of the ${\rm d}M/{\rm d}H$ and $C_p$ data in the 1/9 plateau phase is expected in the fermionic spinon picture, as both are directly related to the spinon density of states (DOS). We note that in the region ($\mu_0 H>10$~T, $T<2$~K) which is our focus in this paper, specific heat contributions from Schottky anomalies and phonons are negligible compared to the intrinsic $C_p$ from the kagome plane, as discussed in the Appendix~\ref{sec_schott}. Note that earlier heat capacity measurements~\cite{Zeng2022,Xu2024, Ray2015} suggest traces of a nuclear Schottky anomaly for $\mu_0 H>10$~T; this may reflect differences between earlier and later sample batches. 

\begin{figure}[!htb]
	
	\begin{center}
		\includegraphics[width= 0.48\textwidth ]{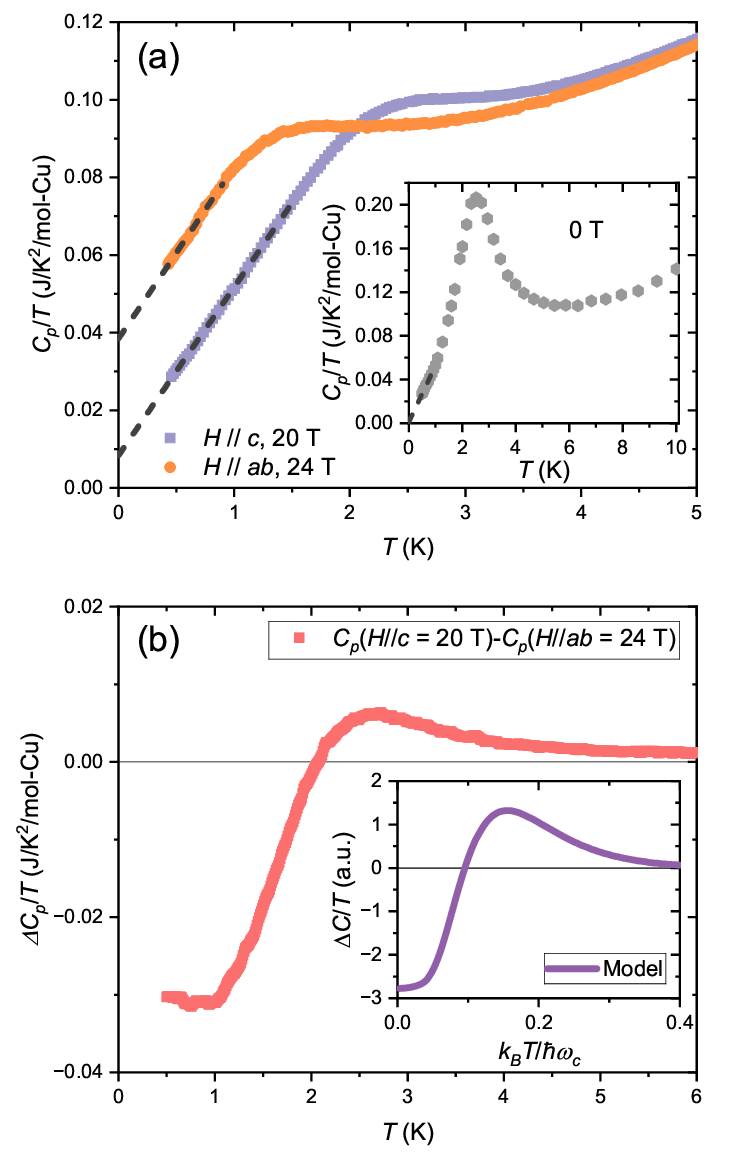}
		\caption{Specific heat data providing evidence for the gapless nature in the 1/9 plateau phase. (a) The raw data of $C_p/T$ vs $T$ are presented down to 0.46 K with $\mu_0 H = 20$~T along $c$ axis (blue dots) and $\mu_0 H = 24$~T in the $ab$ plane (orange dots). The field values are chosen to represent the middle of the plateau after taking into account the $g$-factor anisotropy. The black dashed lines are the linear fittings $C_p/T=\gamma + \beta T$. Note that while the linear slopes are parallel, the finite intercept $\gamma$  strongly depends on the field direction. The inset plot depicts the temperature dependence of specific heat at zero magnetic field, and the black dashed line is a linear fitting with negligible intercept. (b) The temperature dependence of the difference of the curves shown in (a), namely $[C_p(H//c=20~\textup{T})-C_p(H//ab=24~\textup{T})]/T$. The inset plots the result for the same quantity based on the model discussed in Appendix \ref{sec_theory}. } 
		\label{fig2}
	\end{center}
\end{figure}

Zero-field specific-heat data up to 10~K are depicted in the inset of Fig. \ref{fig2}(a). The broad hump around 2.5~K may be explained as the crossover from the paramagnetic to a short-range spin state in QSLs with the temperature scale greatly suppressed from the exchange energy scale $J$ due to frustration, but other explanations are possible~\cite{Radu2005,Yamashita2009, Schnack2018}. As $T$ approaches zero, $C_p/T$ shows linear behavior with a vanishingly small intercept, as indicated by the black dashed linear fit. Our $C_p$ data are in good agreement with the reports in~\cite{Zeng2022, Liu2022}.

\subsection{Fermionic behavior in the vicinity of
the one-ninth plateau}
To investigate the properties of the 1/9 plateau phase, the $T$ dependence of $C_p/T$ within the plateau regions below 5~K with field applied along the $c$-axis and in the $ab$-plane are plotted in Fig. \ref{fig2}(a). We notice that the broad hump shown at 0~T is significantly suppressed in the 1/9 plateau region and no phase transitions are detected at low $T$. This contrasts with the sharp peak feature in specific heat observed near the 1/3 magnetization plateau region reported in some triangular lattices~\cite{Sheng2022}.
Moreover, as $T$ approaches zero, $C_p/T$ shows a linear $T$ behavior with a finite intercept in both directions. The finite intercepts show that the 1/9 plateau phase is gapless with a significant DOS $D(E)$, though the DOS has anisotropy when $H$ is applied in different directions, as already clearly seen in  Fig.~\ref{fig1}(b). The data can be described by a linear fit:  
\begin{equation}
	C_p/T=\gamma + \beta T.
	\label{eq_linear}
\end{equation}
As shown by the black dashed lines in Fig. \ref{fig2}(a), it is clear that the linear slopes are almost parallel, while the intercept $\gamma$ are different in the two directions.  We obtained $\gamma_c=8.0(5)~$mJ/K$^2$/mol-Cu, $\beta_c=43.6(5)~$mJ/K$^3$/mol-Cu for $H \parallel c$, and $\gamma_{ab}=38(1)~$mJ/K$^2$/mol-Cu, $\beta_{ab}=44(1)~$mJ/K$^3$/mol-Cu for $H \parallel ab$. We note that the $\beta$ value is isotropic in the 1/9 plateau phase, while $\gamma$ is highly anisotropic, which suggests that $\gamma$ and $\beta$ terms may have different origins. In a Dirac free fermion,  $C_p/T^2=18 n_D \zeta(3)\pi k_B^3 A_s/ (2\pi \hbar v_D)^2= \beta$, where $n_D$ is the degeneracy of Dirac nodes, $A_s$ is the area of the 2D system, and $v_D$ is the Dirac velocity \cite{Ran2007}. Using $\beta=43.6~$mJ/K$^3$/mol-Cu, we can estimate  $v_D/\sqrt n_D$ to be 1.65 $\times 10^3$ m/s = 10.9 meV$\cdot$\AA. The same quantity was estimated from the approximately linear slope in ${\rm d}M/{\rm d}H$ in Ref. \cite{Zheng2023} to be $(g'/g) \times$ 4.9 meV$\cdot\AA$   where $g'$ is the effective $g$ factor which describes the movement of the down-spin chemical potential in a magnetic field. In Ref. \cite{Zheng2023} $g'/g$ was taken to be $\approx 2$, but there is considerable uncertainty.   Given these uncertainties and the fact that there can be corrections to the free-fermion formulae due to interaction effects, the agreement is reasonable.

We will next focus on $C/T$ for $H \parallel c$ and will return to discuss the case when $H \parallel ab$ and contrast the difference later in the paper.

\begin{figure*}[!htb]
	\begin{center}
		\includegraphics[width=1\textwidth]{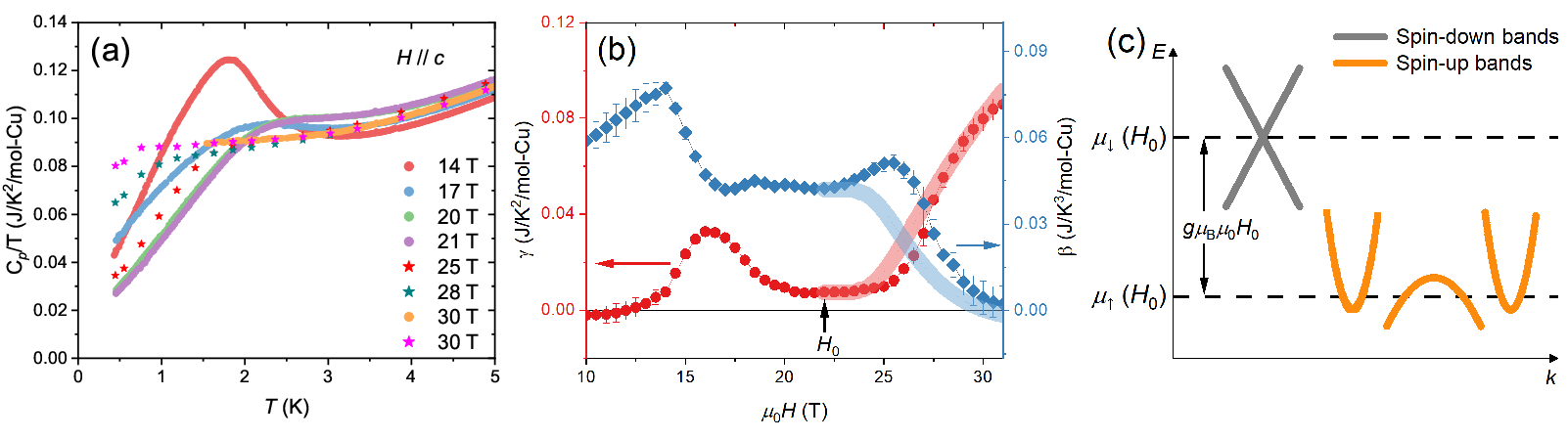}
		\caption{Field evolution of fermionic behavior in the vicinity of 1/9 plateau region. (a) $T$ dependence of $C_p/T$ for different fields with $H \parallel c$. The star-shaped data are cut from Fig.\ref{fig4}(a) at fixed fields. (b)~The field dependence of experimental values of $\gamma$ (red dots) and $\beta$ (blue dots), and the simulated $\gamma$ (thick red curve) and $\beta$ (thick blue curve) obtained from linear fits of $C_p/T$ vs $T$ via Eq. \ref{eq_linear} in the temperature range $0.5\leq T\leq 1.2$~K. The complete experimental and simulated $C_p/T$ vs $T$ data used for fits are shown in Fig.~\ref{FigS_fieldcut} and Fig.~\ref{FigS_Diracsimu}, respectively. The simulation is based on a 2D Dirac spinon (gray bands) centered at the spin-down chemical potential $\mu_{\downarrow}(H_0)=E_0$ combined with a set of particle and hole like bands (orange bands) that cross the spin-up chemical potential $\mu_{\uparrow}(H_0)$, as sketched in (c). The spinon model is described in Section \ref{sec_Dirac}. } 
		\label{fig3}
	\end{center}
\end{figure*}

\begin{figure*}[!htb]
	\begin{center}
		\includegraphics[width= 1\textwidth ]{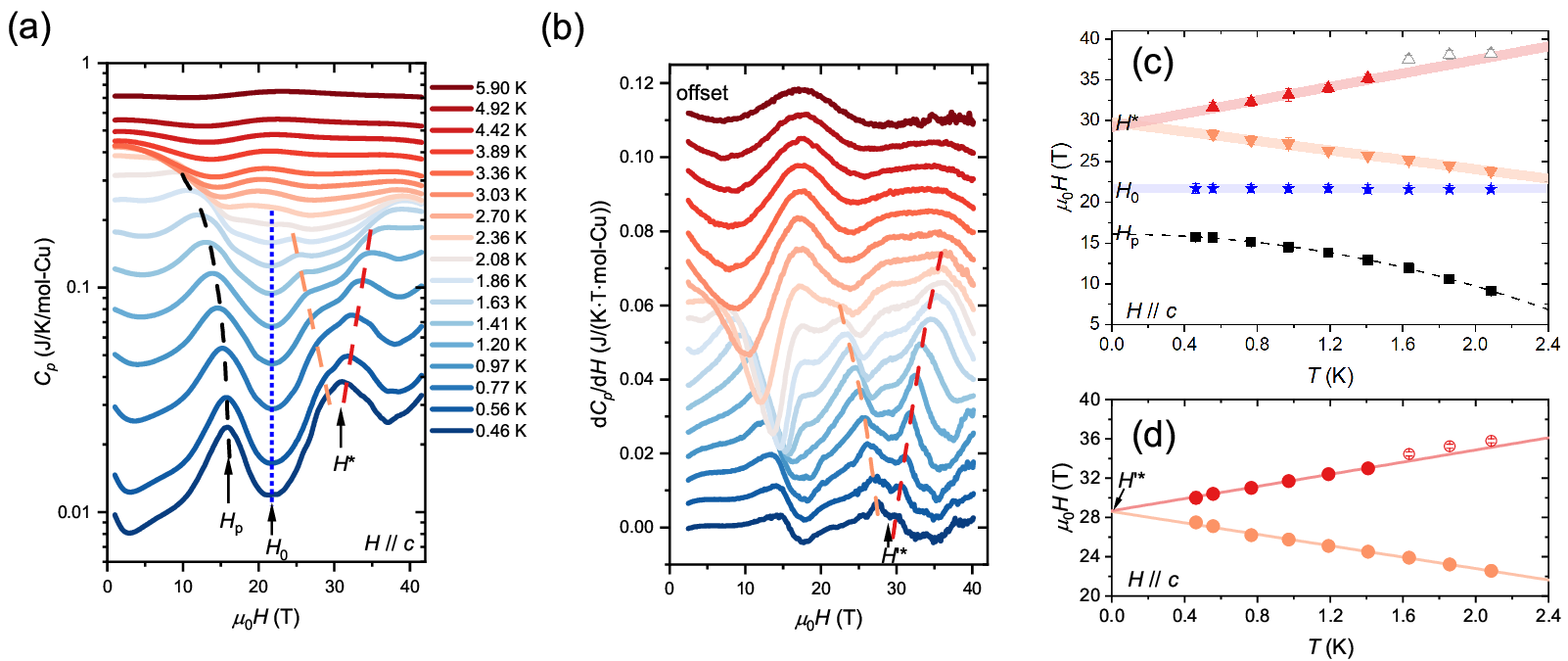}
		\caption{Double-peak structure in the specific heat. (a) the field dependence of $C_p$ at different $T$ with $H$ along the $c$ direction. The black dashed line traces the location of a peak located at $\mu_0 H_p\sim 16$~T at low $T$ and its evolution with increasing $T$, the blue dashed line marks the middle of 1/9 plateau region, and the orange and red dashed lines are guides for tracking the shift of two split peaks at $\mu_0 H^*\sim 30$ T. (b) Field dependence of the derivative of the specific heat ${\rm d}C_p/{\rm d}H$ at different $T$ with a constant offset for clarity. $H^{\prime *}$ indicates the peak-splitting field in ${\rm d}C_p/{\rm d}H$ at 0 K. (c) The dots are the peak or valley locations taken from (a) with the corresponding color codes, and the lines are the fits as described in the main text. (d) The orange and red circles are field locations of two peaks shown by the orange and red dashed lines in (b). The orange and red lines in (d) are two linear fits for the corresponding data points, while the hollow-red circles are excluded from the fits.} 
		\label{fig4}
	\end{center}
\end{figure*}

To gain further insight into the 1/9 plateau phase, the low-temperature dependence of the specific heat over a wide range of magnetic fields ($H \parallel c$) is plotted in Fig. \ref{fig3}(a). Dot-shaped data were obtained with a temperature sweep at a constant field, while star-shaped data were taken from the vertical line-cut in Fig. \ref{fig4}(a) at a fixed field. The overlap of these two different methods at 30~T demonstrates their reliability for further quantitative analysis. A broad hump around 2~K shown in the 14~T data is very similar to the hump observed at zero field in the inset of Fig.~\ref{fig2}(a) and quickly decays as it approaches the 1/9 plateau phase. The intercepts of $C_p$ vs $T$ seem to reach a minimum value inside the 1/9 plateau phase around 21~T. To study the field evolution of $\gamma$ and $\beta$ coefficients, we performed linear fits of the temperature dependence of the specific-heat data based on Eq.~\ref{eq_linear} in the range $0.5\leq T\leq 1.2$~K for different fields. The field dependences of the $\gamma$ (red dots) and $\beta$ (blue dots) coefficients are shown in Fig. \ref{fig3}(b), and the complete $C_p/T$ vs $T$ data sets for the fits are plotted in Fig. \ref{FigS_fieldcut}. When the field is below $\approx 13$~T, the $\gamma$ value is nearly zero, suggesting that this region is an extension of the gapless zero-field state, despite the significant error in $\gamma$ in this field range  ($\approx$~4 mJ/K$^2$/mol-Cu). There is a crossover region from the low-field state to the 1/9 plateau phase indicated by a broad peak in $\gamma$ centered on 16 T. This is associated with our observation in Fig.~\ref{fig1} that the 1/9 magnetization plateau starts around $\mu_0 H=15$~T when $H \parallel c$. Next, inside the central region of the 1/9 plateau between $20-24$~T, both $\gamma$ and $\beta$ coefficients remain almost constant. As the field continues to increase, the $\gamma$ value gradually rises to 86 mJ/K$^2$/mol-Cu at 31~T, while the $\beta$ value shows a small hump around 25.5~T before decreasing to nearly zero at 31~T. The evolution in the vicinity of the 1/9 plateau (20 - 31 T) appears to indicate that the Dirac node is disappearing and a spinon Fermi surface is forming as the chemical potential shifts away from around 22~T, which is consistent with the DSL model under magnetic field proposed in Ref. \cite{Zheng2023}.

\subsection{Dirac spinon model} \label{sec_Dirac}
We therefore attempt to use the model introduced in Ref. \cite{Zheng2023} to explain our specific heat data. In this picture, in the middle of the plateau at field $H_0$, the spin-down chemical potential $\mu_{\downarrow}$ crosses a Dirac spinon band, while the spin-up chemical potential $\mu_{\uparrow}$ crosses electron-like and hole-like bands, forming a spinon semi-metal with total density zero. This is shown in Fig. \ref{fig3}(c). First, the finite $\gamma_c=8$ mJ/K$^2$/mol-Cu around $\mu_0 H_0=22$~T could be attributed to the  bands at $\mu_{\uparrow}$. Let us make the assumption that there is a single hole band, which is heavy, and it is the only one that contributes a $\gamma$ term for $H \parallel c$. The reason for this assumption will be explained later. We adopt the well-known specific heat approximation for free electrons: $C_p=\frac{\pi^2}{3}k_B^2 D_p(E_F) T=\gamma_0 T$, where $D_p(E_F)=n_p m^*/2\pi \hbar$ is a constant DOS for 2D electrons and $n_p$ is the degeneracy which can be set to 1 and $m^*$ is the effective mass of the band. From the former, we obtain an estimate of $m^*\approx11~m_{\rm e}$. Since the specific heat from a parabolic band is independent of $H$, it can be treated as a background constant DOS in Fig. \ref{fig3}(b). Next, we focus on the behavior of the Dirac spinon from spin-down bands. According to Ref.~\cite{Ran2007}, in the low-$T$ limit, we know that $C_p\propto T^2$ at a Dirac node, while $C_p \propto T\cdot  H$ when $k_B T\ll \mu_B \mu_0 H$, where we assume the Dirac node is located at zero-field. However, there is an intermediate field range where $C_p$ can not be described by a simple expression. Thus, we conducted a specific heat simulation on a 2D Dirac node centered at $H_0$ in our case, whose energy dispersion is assumed to be
\begin{equation}
	E=\pm \hbar v_D |\textbf{k}|+E_0
\end{equation}
where $E_0=sg'\mu_B \mu_0 H_0$ describes the energy shift of $s=1/2$ due to Zeeman splitting. We take $g'=g=2.1$ for simplicity; $v_D$ is the only adjustable parameter and can determined by comparison with experiments. This Dirac energy dispersion from spin-down bands is sketched in Fig. \ref{fig3}(c). Then the expression for the DOS of the Dirac spinon is $D_D(E)=\frac{n_D}{2\pi \hbar^2 v_D^2}|E-E_0|$. Next, we can substitute $D_D(E)$ into the expression for the specific heat:
\begin{equation}
	C_p(\mu,T)=\frac{\partial}{\partial T}\int D_D(E)\cdot (E-\mu) \cdot f_{\rm F-D}(E-\mu) {\rm d}E
	\label{eq_C_E}
\end{equation}
Here $\mu$ is the chemical potential and $f_{\rm F-D}(E-\mu)=1/[\textup{exp}(E-\mu)/k_{\rm B}T+1]$ is the Fermi-Dirac distribution function. The simulated $T$-dependence of the specific heat of the Dirac spinon above $H_0$ based on Eq. \ref{eq_C_E} is shown in Fig. \ref{FigS_Diracsimu}, which is consistent with the prediction in Ref. \cite{Ran2007}. The $H$-dependence of $\gamma$ and $\beta$ coefficients obtained from linear fits will be affected by the fitting range of $T$. To compare with the experiments, we chose the same temperature range of $0.5\leq T\leq 1.2$~K as the experimental data to perform the linear fits to the simulated data in Fig. \ref{FigS_Diracsimu}. The fitted $H$-dependence of the $\gamma$ (thick red curve) and $\beta$ (thick blue curve) parameters of the Dirac spinon model are given in Fig. \ref{fig3}(b), by adding the contribution from $\gamma_c$. The resulting value of $v_D$ is essentially the same as that estimated earlier using the linear fits in Fig. \ref{fig2}(a). The Dirac spinon model captures the main features of the experimental data: the flat bottom around $H_0$ in both $\gamma$ and $\beta$, the monotonic increase of spinon Fermi surface, and the disappearance of the Dirac spinon indicated by $\gamma$ and $\beta$ respectively as the field moves above the plateau region. One discrepancy compared to experiments is that no hump feature is observed in simulation for $\beta$ at around 25.5~T. Possible explanations are that the Dirac spinon has a variable (rather than constant) velocity, or that our fitting uncertainties are larger than anticipated. In the above analysis, we focused on the field range $22-31~$T, which could be applied to the low-field regime according to the symmetry of the Dirac node, but the simulations will not be consistent with the experiments below $\approx 19$~T because the DSL model is not applicable in the crossover region.

\subsection{Double-peak structure}
Next we show the full $H$-dependence of $C_p$ for different $T$ for $H \parallel c$ in Fig. \ref{fig4}(a); the corresponding derivative ${\rm d}C_p/{\rm d} H$ is shown in Fig.\ref{fig4}(b). Several remarkable features can be identified in Fig. \ref{fig4}(a) as indicated by colored dashed lines, while their field locations shift with the temperature. To get a better understanding of their evolution, we tracked the locations of these peaks and valleys and plot them in Fig. \ref{fig4}(c). The first feature is the crossover peak at $\mu_0 H_p\approx 16$~T between the zero-field ground state and the 1/9 plateau phase. The $T$-evolution of the peak location is plotted in  Fig.~\ref{fig4}(c) as black squares, which are well fitted by a power law $\mu_0 H-\mu_0 H_p\propto T^q$, where $\mu_0 H_p=16.13(7)$~T and $q=1.98(6)$~T/K$^2$. The interesting quadratic behavior could be related to quantum criticality~\cite{Sachdev2011} separating the low-field state from the plateau state, but this preliminary idea needs to be verified by further detailed experiments. Next is the 1/9 plateau valley centered at $\mu_0 H_0\sim 22$~T, whose location has almost no $T$-dependence as indicated in Fig.~\ref{fig4}(c) by blue stars. Furthermore, a broad peak is seen at $\mu_0 H^*\sim 30$ T in Fig.~\ref{fig4}(a). At first glance, this appears to be a symmetric counterpart of the peak at $H_p$. However, the $T$-dependence of the peak at $H^*$ indicates that it might have a different origin. As $T$ increases, this peak splits into two peaks as tracked by the orange and red dashed lines in Fig.~\ref{fig4}(a) and (b), which can be described by two linear fits with the intercept falling in the same field range as shown in Fig.~\ref{fig4}(c). We note that the red hollow data points in (c) and (d) are excluded from fits because they may be interfered with by another peak in higher field ranges ($>40$ T). Taking the derivative of $C_p$ could make the peak-splitting effect sharper, as displayed in Fig.~\ref{fig4}(b) as orange and red lines. The $T$ dependence of the corresponding peak locations is shown in Fig. \ref{fig4}(d), and the linear fits are carried out using the expression: $\mu_0 H=\mu_0 H^{\prime*} + k T$. The fit results are $\mu_0 H^{\prime*}_1=28.6(3)$~T, $k_1=-2.9(2)$~T/K, and $\mu_0 H^{\prime*}_2=28.7(5)$~T, $k_2=3.1(4)$~T/K for orange and red data points, respectively. The overlapping intercepts and nearly identical slopes are reminiscent of the double-peak structure observed in the specific heat due to a narrow peak in the fermionic DOS~\cite{Yang2023}. To get a better view of the double-peak structure in fermions, we set $(E-\mu)/k_B T=x$ and rewrite Eq. \ref{eq_C_E} as
\begin{equation}
	C_p(x)=k_B^2 T \int D(x)x^2\frac{e^x}{(e^x+1)^2}{\rm d}x.
	\label{eq_C_x}
\end{equation}
Ref.~\cite{Yang2023} called attention  to the double-peak feature in the function $y=x^2\frac{\textup{exp(x)}}{(1+\textup{exp}(x))^2}$.  (The split peaks in $y(x)$ are plotted in Fig.~\ref{FigS_double}(a).) Therefore, a narrow peak in the fermionic DOS $D$ will produce a  linear-in-$T$ splitting in $C_p/T$ as $T$ exceeds the peak width, generating the so-called ``double-peak" structure via Eq. \ref{eq_C_x}. The $T$-dependence of the double-peak locations is shown in Fig.~\ref{FigS_double}(b), which is very similar to what is observed in Fig. \ref{fig4}(c)(d). Thus, by setting $E=sg\mu_B \mu_0 H$ in Fig.~\ref{FigS_double}(b), we obtain the $g$-factor around 30 T for $H$ along the $c$ axis as 2.3(3) or 2.4(2), estimated from $k_1$ and $k_2$ respectively. This remarkable observation provides strong support for the fermionic nature of the excitation. We do not know the origin of the narrow peak, but in the fermionic spinon picture, it could be due to a van Hove singularity away from the Fermi level in one of the spinon bands shown in Fig.~\ref{fig3}(c). 

\subsection{Origin of anisotropy in the one-ninth plateau}
We now return to discuss the difference between the $C/T$ data when $H$ is parallel to the $c$ or lies in the $ab$ plane. The difference is plotted in Fig.~\ref{fig2}(b). The $H$ values along $c$ and $ab$ have been chosen to put us in the middle of the plateau by taking into account the $g$ value anisotropy. We can see that the linear $T$ term almost cancels, resulting in a virtually constant value below $1.2$~K, suggesting a gap-like behavior. In the fit given by Eq.~\ref{eq_linear}, $\beta$ is isotropic while $\gamma$ depends on the field direction and is described by $\gamma_{ab}$ and $\gamma_c$. Apparently, $C/T$ is suppressed with $H \parallel c$ in a temperature-dependent way, resulting in a $\gamma_c$ at low temperature that is much smaller (about 1/5) compared with $\gamma_{ab}$. We emphasize that this kind of strong anisotropy is rather surprising. In a spin 1/2 Heisenberg model, the heat capacity is strictly isotropic. In a magnetically ordered state, the magnon may show an anisotropic gap in the presence of spin-orbit coupling (SOC), such as the Dzyaloshinskii–Moriya (DM) term. However, in these examples, there are no residual $\gamma$, let alone a $\gamma$ value that depends on the orientation of the magnetic field. The dependence on $H_c$, the field component along the $c$-axis, suggests that an orbital degree of freedom is at play. The orbital effect is central to the picture proposed in our earlier paper to explain the quantum oscillations, which were demonstrated to depend on $H_c$.~\cite{Zheng2023} This leads us to propose an extension of our earlier model to give an account of this interesting observation.

Recall that at the middle of the plateau at $H=H_0$ the spin-down chemical potential $\mu_{\downarrow}$ crosses a Dirac spinon band, while the spin-up chemical potential $\mu_{\uparrow}$ crosses particle-like and hole-like bands, forming a spinon semi-metal. This is shown in Fig. \ref{fig3}(c). In Ref.~\cite{Zheng2023}, the focus was on the Landau levels formed in the Dirac band when $H_c$ is nonzero. The idea is that due to the DM term, an external magnetic field along $c$ produces a gauge magnetic field ${\cal B}$ that acts on the spinon in an analogous way to a conventional magnetic field~\cite{Kang2024},  forming Landau levels which give rise to quantum oscillations. It is natural to extend the notion of Landau quantization to the bands that cross $\mu_{\uparrow}$. For concreteness, we assume a single heavy hole band at the zone center with effective mass $m_{\rm h}=11 m_{\rm e}$ to account for $\gamma_c$ as discussed earlier. We assume 6 lighter particle-like bands with mass $m_p \approx m_h/2$ to give an additional contribution to $\gamma_{ab}$. (Owing to the three-fold symmetry, the bands away from the zone center come in multiples of 3's, so the assumption of 6 particle-like bands is not unreasonable.) For $H \parallel c$ the Landau bands are formed in these lighter particle-like bands. (For simplicity, we assume that the Landau level spacing in the hole-like band has a negligible effect on $C/T$ at the lowest experimentally available $T$ due to the heavier mass.) If the chemical potential is pinned between Landau levels, the contributions from the particle-like bands will be suppressed and show a gap-like behavior on a temperature scale given by the cyclotron frequency $\omega_c$. A detailed analysis is given in Appendix \ref{sec_theory}, and the theoretical curve is shown in the inset to Fig.~\ref{fig2}(b). 

Note the activation gap at low temperatures and the appearance o of a positive hump before $C/T$ saturates to a constant value at high temperatures. The origin of this hump is the same as the origin of the split peak shown in Fig. \ref{fig4}(c),(d). It is associated with the double peak in Eq.~\ref{eq_C_x} due to a narrow peak in the DOS, which splits into two as $T$ is increased. In this case, the Landau level closest to the Fermi level is the narrow peak that splits, and its tail gives rise to the bump at the chemical potential. We should emphasize that the picture of Fermi level pinning half-way between Landau levels is very different from the conventional picture of Landau levels, where the Landau levels move across the Fermi level to give rise to quantum oscillations. In our case, there are no quantum oscillations from these bands because the chemical potential is pinned. The rationale for the Landau-level pinning is that the spinon bands are the results of the solution of a self-consistent set of mean-field equations to minimize the free energy. There is a gain in free energy by placing the chemical potential in the middle of the gap, as is common in any mean field theory. See Appendix \ref{sec_theory} for further discussions.

\section{Discussion}
Finally, we compare our observations with specific-heat results in other QSL candidates. A finite $\gamma$ value has been reported in different frustrated systems. Notably, the famous organic materials $\kappa-$(BEDT-TTF)$_2$Cu$_2$(CN)$_3$ \cite{Yamashita2008} and EtMe$_3$Sb[Pd(dmit)$_2$]$_2$ \cite{Yamashita2011} provided early evidence of spinon Fermi surface ground states. In another KHA material, herbertsmithite, a $\gamma$ value of 50~mJ/K$^2$/mol has also been observed, but no 1/9 magnetization plateau has been reported so far, and the field dependence of the specific heat at high fields is featureless \cite{Barthelemy2022, Han2014}. The field-dependent DOS and Dirac velocity have scarcely been studied. One exception is the quasi-linear field dependence of the Dirac velocity obtained from $\alpha$ originating from the Majorana-fermions in the Kitaev magnet $\alpha$-RuCl$_3$~\cite{Imamura2024}. The simultaneous observation of a constant $\gamma$ term and a $\beta$ term linear in $T$ has not been reported before the current work. Furthermore, the 1/9 plateau phase exhibits specific-heat characteristics that are entirely different from those of the trivial 1/3 plateau. For instance, the sharp $\lambda$-like peak feature in the $T$ dependence of $C_p$ around the gapped 1/3 plateau phase boundary in some triangular lattices, like Cs$_2$CuCl$_4$~\cite{Radu2005} and Na$_2$BaCo(PO$_4$)$_2$~\cite{Sheng2022}, is a signature of the transition into magnetically ordered states. Conversely, no sharp peak has been observed in the $T$ dependence of $C_p$ down to 0.46~K within the gapless 1/9 plateau phase as shown in Fig. \ref{fig2}(a). This difference strongly suggests that the 1/9 plateau phase could be an exotic spin-liquid plateau induced by the magnetic field~\cite{Nishimoto2013}. 

\section{Conclusions}
In summary, we observed the unconventional 1/9 plateau in both the magnetization and specific heat in YCOB. The temperature dependence of the specific heat provides evidence that the 1/9 plateau is gapless with a finite DOS. Further field dependent analysis indicates there could be a DSL in the 1/9 plateau phase centered at 22~T, which gradually evolves into a spinon Fermi surface at around 30~T. The double-peak structure observed at 30~T gives further evidence for the Fermionic excitations. The strong anisotropy of the $\gamma$ term shows that orbital effects may be at play. Our results provide direct low-energy excitation information to understand the 1/9 plateau phase, providing evidence for an exotic DSL state associated with this plateau. These discoveries could be a significant step for the search of QSL and the study of quantum entangled states. 

{\it Acknowledgement}
The work at the University of Michigan is primarily supported by the National Science Foundation under Award No.DMR-2317618 (thermodynamic measurements) to Kuan-Wen Chen, Dechen Zhang, Guoxin Zheng, Aaron Chan, Yuan Zhu, Kaila Jenkins, and Lu Li. The magnetization measurements at the University of Michigan are supported by the Department of Energy under Award No. DE-SC0020184. A portion of this work was performed at the National High Magnetic Field Laboratory (NHMFL), which is supported by National Science Foundation Cooperative Agreement Nos. DMR-1644779 and DMR-2128556 and the Department of Energy (DOE). J.S. acknowledges support from the DOE BES program “Science at 100 T,” which permitted the design and construction of much of the specialized equipment used in the high-field studies. The work at IOP China is supported for the crystal growth, by the National Key Research and Development Program of China (Grants 2022YFA1403400, No. 2021YFA1400401), the K. C. Wong Education Foundation (Grants No. GJTD-2020-01), the Strategic Priority Research Program (B) of the Chinese Academy of Sciences (Grants No. XDB33000000). The experiment in NHMFL is funded in part by a QuantEmX grant from ICAM and the Gordon and Betty Moore Foundation through Grant No. GBMF5305 to Kuan-Wen Chen, Dechen Zhang, Guoxin Zheng, Aaron Chan, Yuan Zhu, and Kaila Jenkins. P.L. acknowledges the support by DOE office of Basic Sciences Grant No. DE-FG02-03ER46076 (theory).

\appendix
\begin{figure*}[!htb]
	
	\begin{center}
		\includegraphics[width= 1\textwidth ]{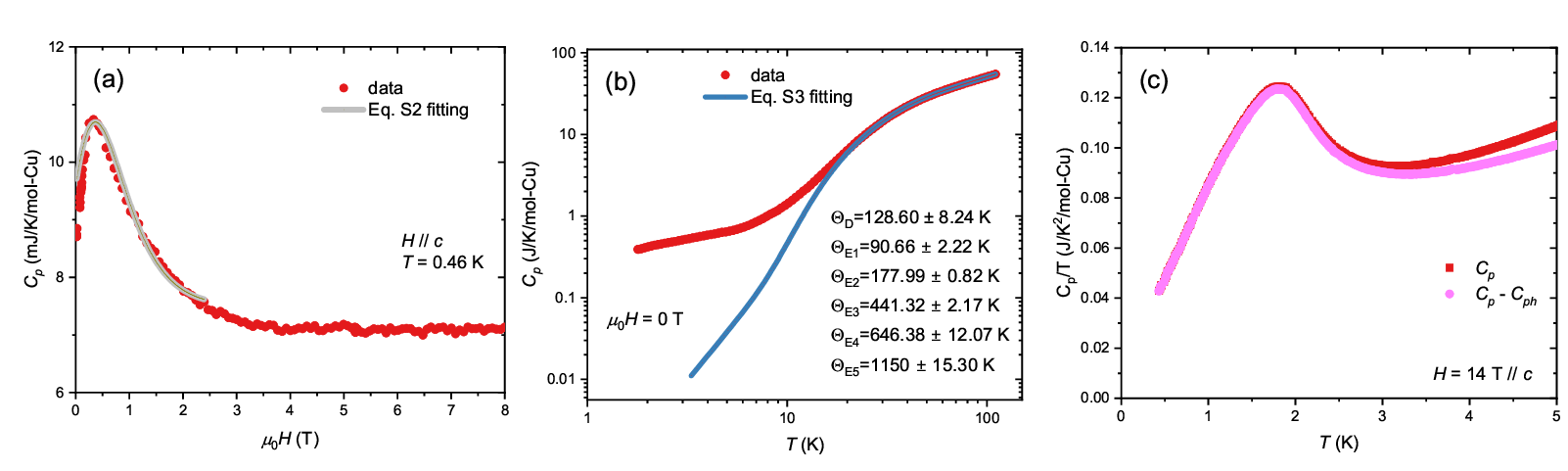}
		\caption{Specific heat contributions from Schottky and phonon terms. (a) The red dots represent the $C_p$ data taken from Fig. 1(b) in the main text for $H\parallel c$ after subtracting a linear background. The gray curve is the Schottky contribution fitted using Eq.~\ref{eq_C_sc} in the range of $0-2.5$~T. (b) The red dots are the raw $C_p$ data taken from 1.8~K to 110~K at 0~T in the PPMS. The blue curve is the phonon specific heat fitted using Eq.~\ref{eq_phonon}. The best-fit parameters are listed in the figure. (c) The red dots are the $\mu_0 H=14$~T $C_p$ data presented in Fig.~\ref{fig3}(a) in the main text. The blue dots are the result of $C_p-C_{ph}$, i.e. a phonon contribution is subtracted from the red dots obtained from the fits in (b).  } 
		\label{FigS_Schott}
	\end{center}
\end{figure*}

\section{Specific heat due to Schottky and phonon contributions} \label{sec_schott}
To analyze the total specific heat, we used the following expression:
\begin{equation}
	C_p=C_{\rm{sc}}+C_{\rm{ph}}+C_{\rm{ka}}
\end{equation}
where $C_{\rm{sc}}$ is the Schottky-like contribution arising from the localized excitations, $C_{\rm{ph}}$ is the conventional phonon contribution, and $C_{\rm{ka}}$ is the specific heat originating from the kagome plane. As shown in Fig. \ref{fig1}(b) in the main text, a Schottky-like anomaly was observed at a low field range ($< 0.3$~T) in both directions, which can be fitted by a two-level Schottky model:
\begin{equation}
	C_{\rm{sc}}=f\frac{N_Ak_{\rm B}\Delta^2 e^{\Delta/T}}{T^2(1+e^{\Delta/T})^2}
    \label{eq_C_sc}
\end{equation}
where $f$ is the fraction of orphan spins, $\Delta$ is the energy gap following $\Delta=g\mu_B\mu_0 H/k_{\rm B}+\Delta_0$ with a field-independent gap $\Delta_0$. The fitted result is shown in Fig.~\ref{FigS_Schott}(a) using parameters $f=0.087$\%, $g=2$, and $\Delta_0=0.6$~K. A linear density of states (DOS) contribution from the kagome plane has been subtracted from raw $C_p$ data in Fig.~\ref{FigS_Schott}(a) to fit the Schottky anomaly correctly. $C_{\rm{sc}}$ quickly decays and becomes negligible when the field is higher than 10~T. To estimate the contribution of $C_{\rm{ph}}$, we applied a Debye-Einstein function~\cite{Liu2022} to fit $C_p$ vs $T$ from 30~K to 110~K:
\begin{equation}
	3C_{\rm{ph}}=\frac{9RT^3}{\Theta_{\rm D}^3}\int_{0}^{\Theta_{\rm D}/T} \frac{\xi ^4 e^\xi}{(e^\xi-1)^2} {\rm d}0\xi +\frac{R}{T^2}\sum_{n=1}^{5} \frac{w_n\Theta_{\rm En}^2 e^{\Theta_{\rm En}/T}}{(e^{\Theta_{\rm En}/T}-1)^2}
	\label{eq_phonon}
\end{equation}
where $\Theta_{\rm D}$ and $\Theta_{\rm En}$ are fitting parameters, and $w_n$ are the weights for different $\Theta_{\rm En}$. The fitting result and fitted parameters are shown in Fig.~\ref{FigS_Schott}(b). The fitted $C_{\rm{ph}}$ was extended to low $T$ and compared with the total specific heat in Fig.~\ref{FigS_Schott}(c), which shows that $C_{\rm{ph}}$ is negligible when $T$ is below 2 K. Therefore, we conclude that in the region ($\mu_0 H>10$~T, $T<2$~K) that we focus on in this study, it should be safe to use the estimate $C_p \approx C_{\rm{ka}}$. High fields naturally separate the intrinsic specific-heat contributions induced by kagome frustrations from extrinsic localized excitation parts produced by orphan spins or band-randomness \cite{Liu2022,Kawamura2014,Shivaram2024} which have introduced controversial results in the ground state at zero field \cite{Zeng2022,Liu2022,Hong2022}.

\begin{figure}[!htb]
	
	\begin{center}
		\includegraphics[width= 0.5\textwidth ]{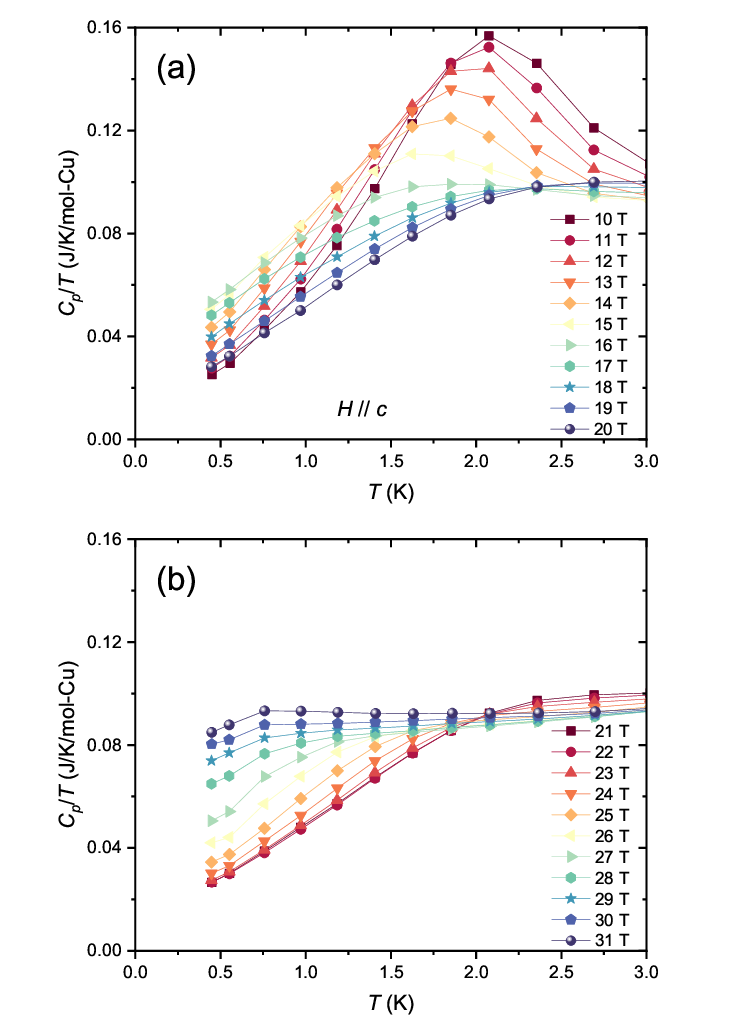}
		\caption{The line-cut temperature dependence of $C_p/T$  under different magnetic fields taken from the data in Fig. \ref{fig4}(a) in the main text. The $10-20$~T curves are shown in (a), and the $21-31$~T curves are plotted in (b).} 
		\label{FigS_fieldcut}
	\end{center}
\end{figure}

\begin{figure}[!htb]
	
	\begin{center}
		\includegraphics[width= 0.5\textwidth ]{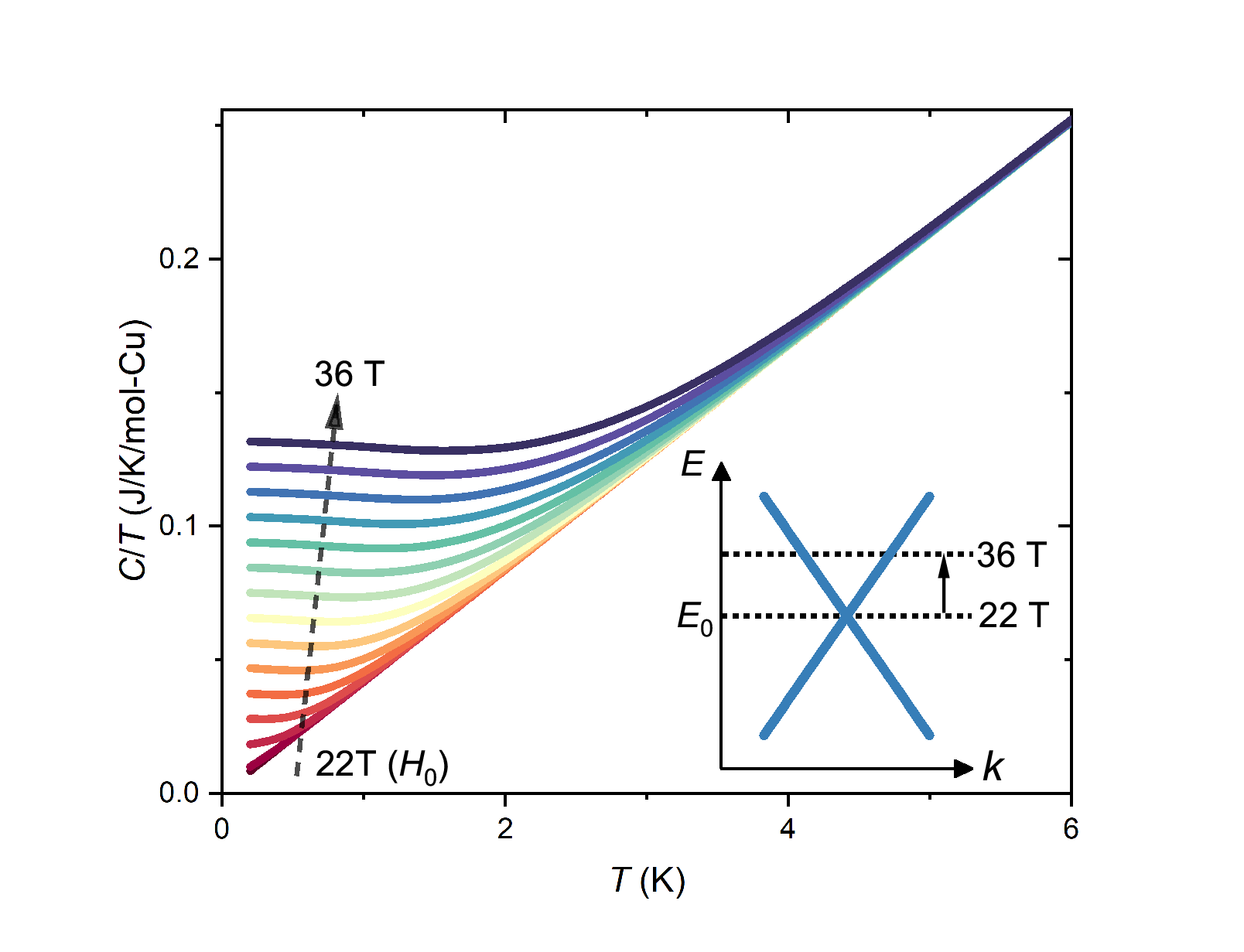}
		\caption{The simulated $C/T$ vs $T$ of a Dirac spinon coming from spin-down bands under different magnetic fields, without considering the Landau Levels. The band structure of the model is shown in the inset, which has a Dirac linear energy dispersion with the crossing point at $E(H_0)$ where $\mu_0 H_0=22$~T. When the magnetic field increases from 22~T to 36~T, the spinon Fermi surface will grow monotonically due to the Zeeman effect. } 
		\label{FigS_Diracsimu}
	\end{center}
\end{figure}

\begin{figure}[!htb]
	
	\begin{center}
		\includegraphics[width= 0.48\textwidth ]{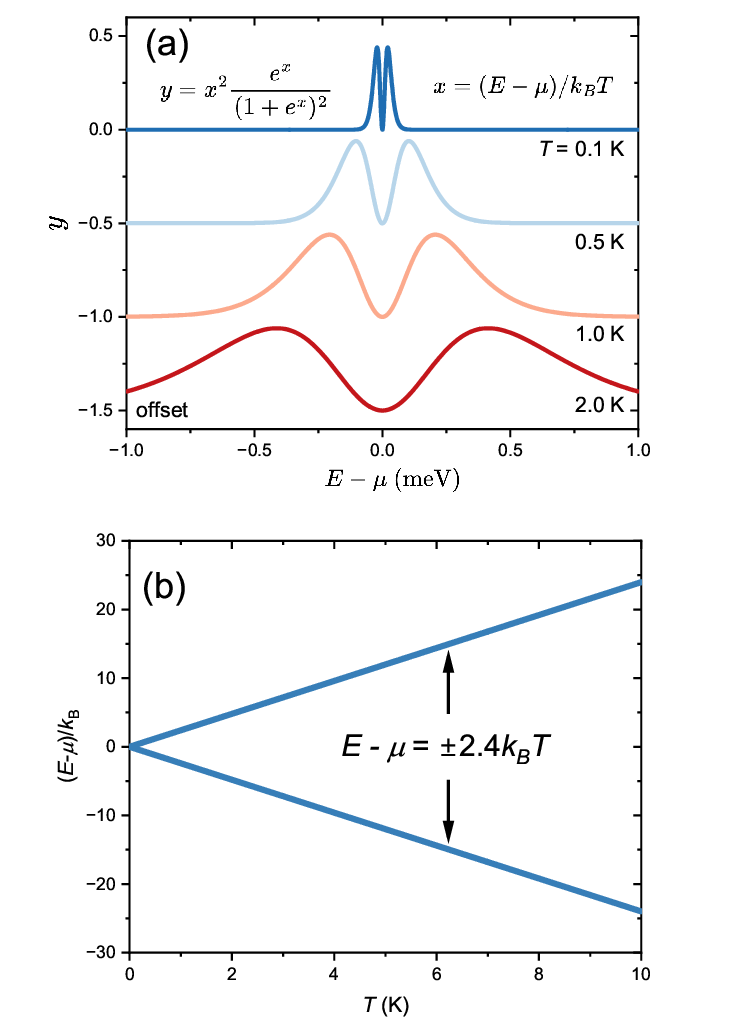}
		\caption{The calculations of the double-peak structure. (a) The temperature evolution of $y=x^2 \textup{exp}(x)/(1+\textup{exp}(x))^2$, where $x=(E-\mu)/k_{\rm B} T$, and $\mu$ is the chemical potential. (b) The temperature dependence of the locations of two peaks in (a), which can be described by the linear equations $E-\mu=\pm 2.4k_{\rm B} T$. } 
		\label{FigS_double}
	\end{center}
\end{figure}

\section{Theoretical model for anisotropy} \label{sec_theory}
In this section,  we present a model that ascribes the difference between $C/T$ for magnetic field $H$ applied along $c$- and $ab$-axis to the Landau level gap that opens up in one of the spin-up spinon bands. Recall that at the middle of the 1/9 plateau at $H=H_0$, the spin-up bands form a ``semi-metal'' consisting of particle and hole bands while the spin-down band is assumed to obey a Dirac spectrum. First, we focus on the spin-up bands. For concreteness, we assume a single parabolic hole band at the zone center with mass $m_{\rm h}$ and six parabolic particle bands with mass $m_{\rm p}$. Owing to the 3-fold symmetry, the assumption of  6 bands is reasonable if they are located away from the zone center. We assume that $m_{\rm h}=2 m_{\rm p}$ so that the particle bands contribute three times as much as the hole band to $C/T$ at low temperatures. In the model presented below, the particle bands are gapped when $H$ is along $\textrm{c}$, thereby explaining the roughly factor of 4 anisotropy in the $\gamma$ term. The idea is that  Landau levels are formed in the presence of $H_c$. We assume that at the lowest temperature achieved in the experiment, the effect of quantization of the hole band is negligible due to its heavier mass, and concentrate on the lighter particle band.  Landau levels are formed with energy $\omega_{\rm c} = \frac{e{\cal B}}{m_{\rm pc}}$ where ${\cal B}$ is the gauge magnetic field given by ${\cal B}=\alpha B \cos \theta$, where $B\approx \mu_0 H$, $\alpha$ is a constant~\cite{Zheng2023, Kang2024}, and $\theta$ is defined as the angle between $c$ and $ab$ axes. We make a key assumption that the chemical potential is located in the middle of Landau levels, even if the Landau level spacing is modified by changing the $c$ component of the magnetic field $B \cos \theta$. This is very different from the usual picture of Landau levels, where the chemical potential is approximately fixed, and the Landau levels move across the Fermi energy as $\cos \theta$ is varied, giving rise to quantum oscillations. With our assumption, the Landau levels do not move across the chemical potential, and there is no quantum oscillation for low energy excitations near the Fermi energy. Instead, it is the bottom of the parabolic band that changes as the number of occupied Landau levels is changed as a function of $\theta$, and the density exhibits quantum oscillations. This scenario is possible for the spinons because the spinon band structure, including the location of the band bottom, is determined by a self-consistent solution of some mean field equations. Keeping the chemical potential midway between Landau levels lowers the kinetic energy to take advantage of the gap. The density is allowed to vary because the up-spin spinon forms a semi-metal, consisting of particle and hole bands, and the constraint is that the total density is fixed. Our consideration is for one of the particle or hole bands, and its density is allowed to vary. This assumption is necessary to explain the data because in the standard picture, $C/T$ will show oscillations as a function of $\theta$ at low temperatures, which is not seen in the experiment. 
Starting with this model, we make a further assumption that the number of occupied Landau levels is large enough and the temperature is low enough so that we can extend the summation of the occupied Landau levels to negative infinity. In this case, The heat capacity is given by 
\begin{equation}
\label{eq:heat-capacity-integral}
C = \int_{-\infty}^{\infty} {\rm d}E \, (E-\mu) \, D(E; \tau) \frac{\partial f (E - \mu; T)}{\partial T} , 
\end{equation}
where $D(E; \tau)$ is the density of states and $f(E - \mu; T) = \frac{1}{e^{(E - \mu)/k_{\rm B}T} + 1}$ is the Fermi-Dirac distribution function. We will set $\mu=0$ from now on. We assume that the density of states has a disorder broadening of the form: 
\begin{equation}
\label{eq:heat-capacity-DOS}
D(E; \tau) = \sum_{n=-\infty}^{\infty} \frac{1}{2\pi \tau} \frac{\cos \theta}{(E - E_n \cos \theta)^2 + \big( \frac{1}{2\tau} \big)^2} , 
\end{equation}
where $\tau$ is the characteristic inverse energy scale with $\frac{1}{\tau}$ being a full width at half maximum and $E_n = \big( n + \frac{1}{2} \big) \hbar \omega_{\rm c}$ are the Landau level energies of the spin-up spinon. $\cos \theta$ in the numerator in the density of states $D(E; \tau)$ is the degeneracy factor of each Landau level.

\subsection{Numerical Results for Spin-up Spinon}
In the following, we present the heat capacity calculation mainly for the spin-up spinon, and then briefly discuss the spin-down spinon. We numerically evaluate the integral expression of the heat capacity Eq.~\eqref{eq:heat-capacity-integral} as a function of temperature and angle. 

\begin{figure}[!htb]
	\begin{center}
		\includegraphics[width=0.48\textwidth]{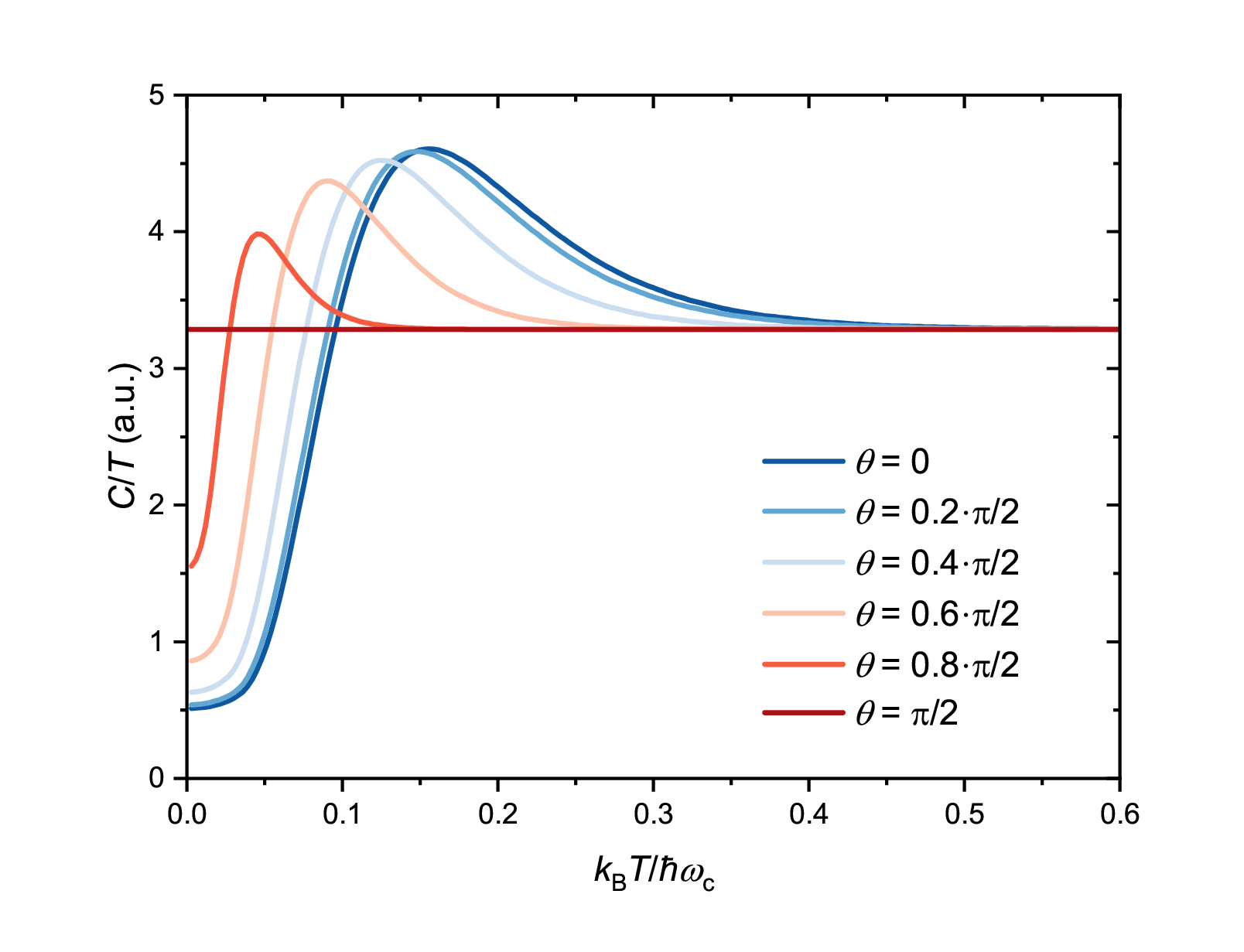}
		\caption{Heat capacity over temperature as a function of temperature for different angles. We set $\tau = 10/\omega_c$ in our calculation. As we increase the angle $\theta$ from $0$, the curve becomes more flat and eventually becomes a constant function in the $\theta \to \frac{\pi}{2}$ (in-plane magnetic field) limit.} 
		\label{FigS_heat_capacity_T_theory}
	\end{center}
\end{figure}

The spin-up spinons form ordinary Landau levels upon applying an external field. In Fig.~\ref{FigS_heat_capacity_T_theory}, we plot the heat capacity over the temperature ($C/T$) as a function of the temperature. The temperature is measured in the unit of the Landau level spacing $\hbar \omega_{\rm c}$ when $B$-field is along $\textrm{c}$-axis, i.e., when $\theta = 0$. As we tilt the angle of the magnetic field from $\theta = 0$ to $\theta = \frac{\pi}{2}$, $C/T$ becomes flatter and flatter as a function of $T$, and eventually becomes a constant function in the limit of in-plane magnetic field ($\theta = \frac{\pi}{2}$). Thus, if we subtract $\theta = 0$ curve from $\theta = \frac{\pi}{2}$ curve, it shows a gap-like behavior that reproduces the overall shape of the curve from the experimental data (Fig. \ref{fig2}(b) in the main text). Notice that there is a broad peak in $C/T$ just above the gap. The origin of this peak is due to the double-peak structure in Eq. \ref{eq_C_x} in the main text, which leads to the splitting of a narrow peak at temperatures larger than its width after performing the integration. At temperatures high compared with the Landau level width, the  Landau level that is closest to the Fermi level splits, and its tail at zero energy gives rise to a peak in $C/T$. The peak in the theory appears to be more prominent than the peak in the data, which is subject to uncertainty due to the subtraction procedure using data at different $B$ fields to compensate for the $g$-factor anisotropy.

\begin{figure}[!htb]
	\begin{center}
		\includegraphics[width=0.48\textwidth]{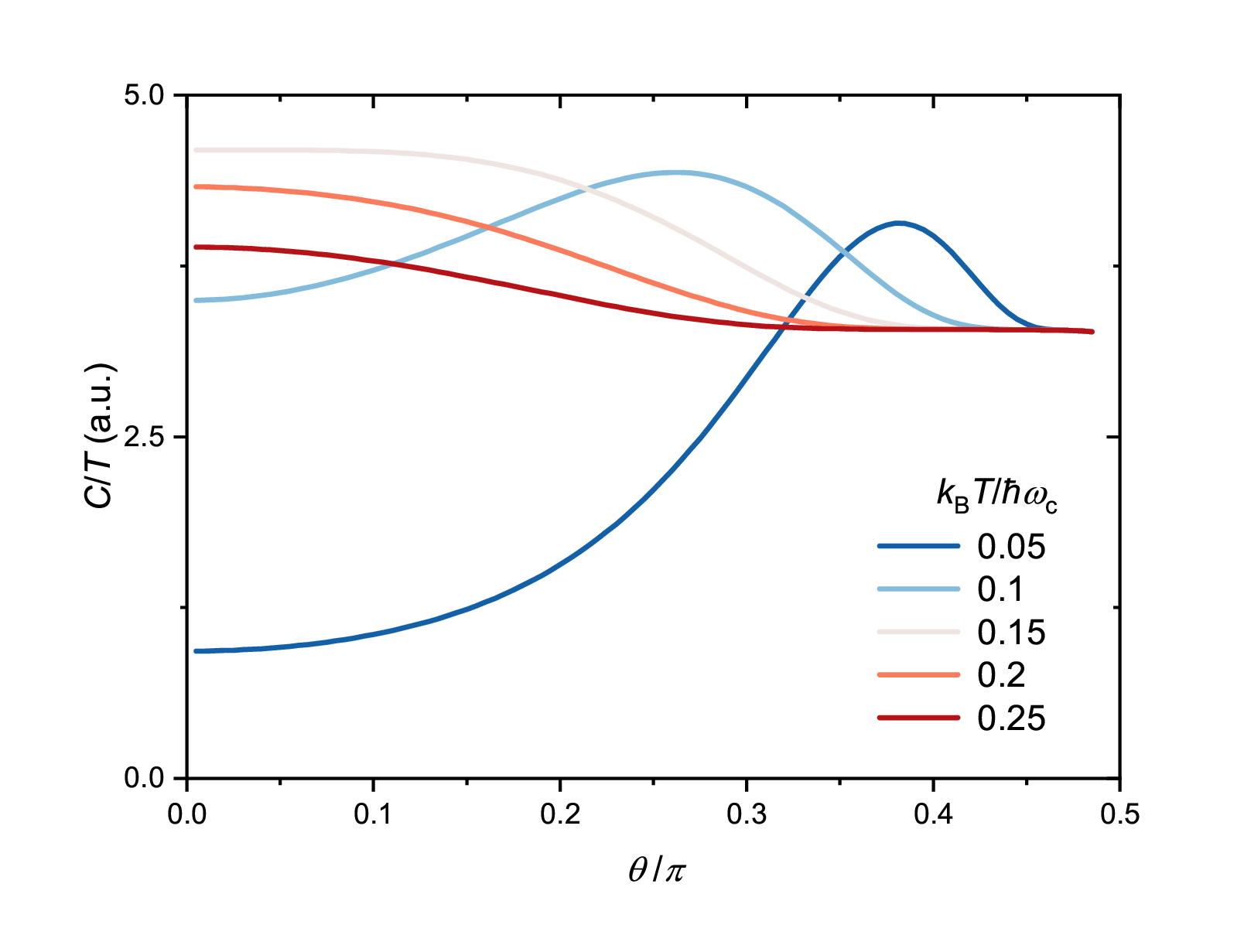}
		\caption{Heat capacity over temperature ($C/T$) as a function of angle ($\theta$) for different temperatures. We set $\tau = 10/\omega_c$ in our calculation.}  
		\label{FigS_heat_capacity_theta_theory}
	\end{center}
\end{figure}

Next, we compute $C/T$ as a function of angle $\theta$ for various values of temperature, which is presented in Fig.~\ref{FigS_heat_capacity_theta_theory}. We used $k_{\rm B} T/\hbar \omega_{\rm c}=0.05$ as our lowest temperature value, which corresponds to the temperature at which $\theta = 0$ curve in our Fig.~\ref{FigS_heat_capacity_T_theory} shows a crossover behavior from the initial low temperature plateau to increasing behavior. $k_{\rm B} T/\hbar \omega_{\rm c}=0.05$ corresponds to 1 K in the experiment. As expected, the $C/T$ curve becomes more flat as we increase the temperature.

\subsection{Numerical Results for Spin-down Spinon}
\begin{figure}[!htb]
	\begin{center}
		\includegraphics[width=0.48\textwidth]{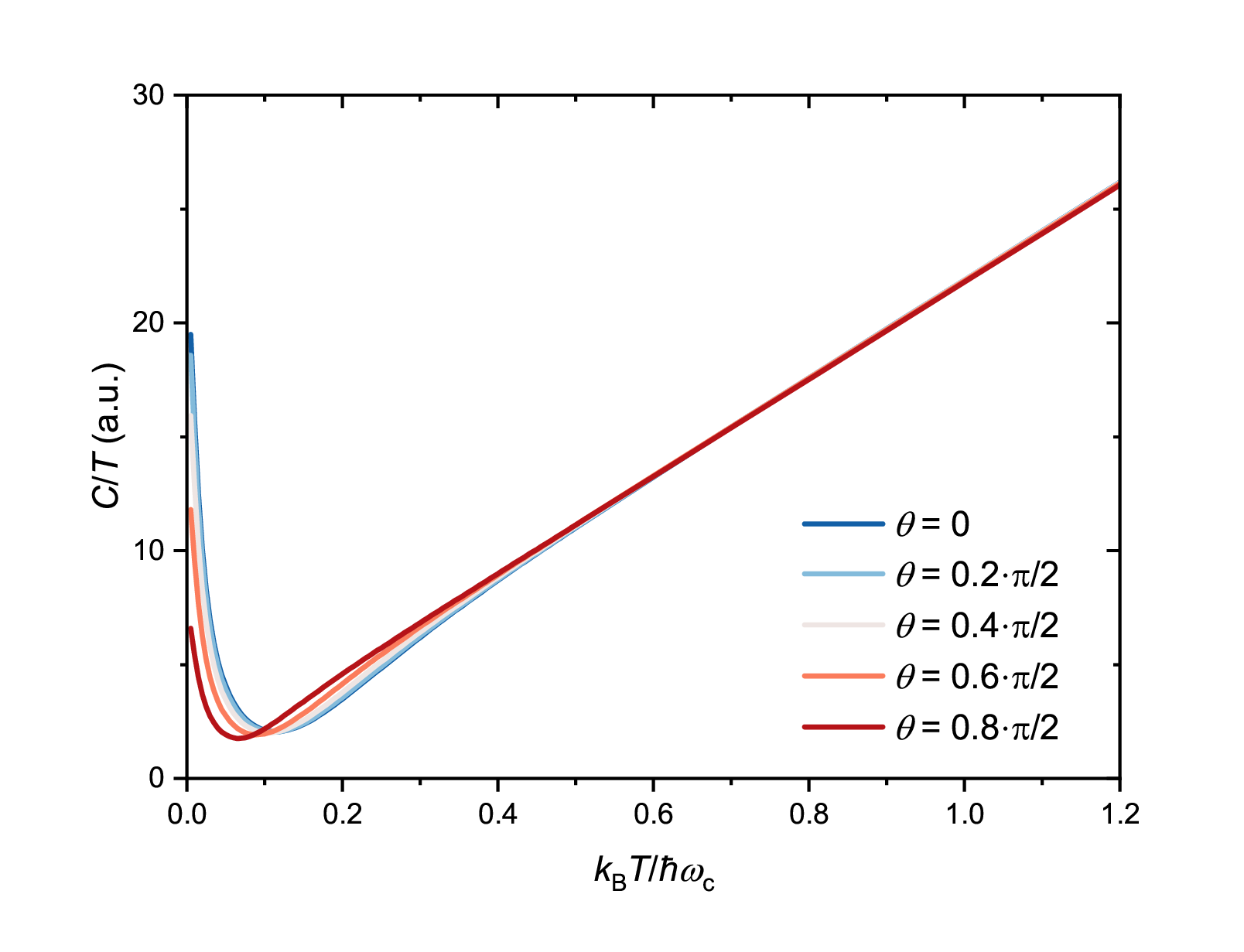}
		\caption{$C/T$ as a function of the temperature for various angles in the spin-down spinon (Dirac fermion) case, after including the effect of Landau levels. The relaxation time $\tau$ is chosen to be 10. All energy scales are measured in units of the spacing between the $n=0$ and $n=1$ Landau levels.}
		\label{FigS_heat_capacity_dirac_theory}
	\end{center}
\end{figure}

In this section, we address the question of whether the spin-down spinon will also show significant anisotropy. 
The spin-down spinon is assumed to follow a Dirac spectrum and thus the Landau level energy $E_n$ is equal to $\sqrt{n}$, where the energy is measured with respect to the Landau level energy gap between $n=1$ and $n=0$. In this case, we can use the particle-hole symmetry to pin the chemical potential $\mu = 0$. Using Eqs.~\ref{eq:heat-capacity-integral} and \ref{eq:heat-capacity-DOS}, the temperature dependence of $C/T$ is shown in Fig.~\ref{FigS_heat_capacity_dirac_theory}. Notice that, unlike the parabolic spectrum, there is a zero mode in the Dirac case, which gives rise to an upturn in $C/T$ for $k_B T/\hbar \omega_c$ approximately less than 0.07. Above this scale, the anisotropy is small. In the experiment, this upturn is not visible, presumably because the temperature is not low enough. Alternatively, there may be a small gap in the Dirac spectrum, or the zero mode is absent for some reason that is not understood. We note that in Fig. \ref{FigS_Diracsimu}, the effect of Landau quantization was not included. This is why the upturn at the very low temperature in Fig. \ref{FigS_heat_capacity_dirac_theory} is not visible there.

\end{document}